\documentstyle[prl,twocolumn,epsfig,aps]{revtex}

\begin{document}

\draft

\title{Interfaces (and Regional Congruence?) in Spin Glasses}  
\author{C.M.~Newman}
\address{Courant Inst.~of Math.~Sciences,
New York Univ., New York, NY 10012}
\author{D.L.~Stein}
\address{Departments of Physics and Mathematics, University of Arizona,
Tucson, AZ 85721}
\maketitle

\begin{abstract} 
We present a general theorem restricting properties of
interfaces between thermodynamic states and apply it 
to the spin glass excitations observed
numerically by Krzakala-Martin and Palassini-Young in spatial
dimensions $d=3,4$. We show that such excitations, with interface
dimension $d_s<d$, {\it cannot\/} yield regionally congruent
thermodynamic states. More generally,  
zero density interfaces of translation-covariant
excitations cannot be pinned (by the disorder) in any $d$ but rather
must deflect to infinity in the thermodynamic limit.  Additional
consequences concerning regional congruence in spin glasses and other
systems are discussed.

\end{abstract}


%\narrowtext

Recent numerical studies \cite{KM00,PY00,HKM,KPY} on
finite-dimensional spin glasses observed unanticipated low-energy
excitations that have generated considerable interest
\cite{Middleton00,MP00a,MP00b,HHD}.  The models examined were
nearest-neighbor Ising spin glasses with the Edwards-Anderson
\cite{EA} Hamiltonian
\begin{equation}
\label{eq:Hamiltonian}
{\cal H}=-\sum_{<x,y>} J_{xy} \sigma_x\sigma_y \ ,
\end{equation}
where the couplings $J_{xy}$ are independently chosen
from a Gaussian distribution with mean zero
and variance one, the sum is over only nearest neighbors on the
$d$-dimensional cubic lattice, and the spins $\sigma_z=\pm 1$.

These studies examined the ground states $\pm \sigma^L$ and their
low-energy excitations corresponding to the coupling realization ${\cal
J}^L$ inside cubes $\Lambda_L$ of side $L$, centered at the origin, with
periodic boundary conditions.  Krzakala and Martin (KM) \cite{KM00},
working in $d=3$, forced a random pair of spins, $\sigma_z,\sigma_{z'}$, to
assume a relative orientation opposite to the one in $\sigma^L$, and then
relaxed the rest of the spins to a new lowest-energy configuration.  This
ensures that at least some bonds in the excited spin configuration,
$\sigma'^L$, must be changed (i.e., satisfied $\leftrightarrow$
unsatisfied) from $\sigma^L$.  It also ensures that the energy of
$\sigma'^L$ is no more than $O(1)$ above that of $\sigma_L$, regardless of
$L$.  Interestingly, rather than simply generating local droplet flips, the
KM procedure was observed to yield large-scale (i.e., of lengthscale
$l=O(L)$) excitations $\sigma'^L$, with $O(L^d)$ spins flipped from
$\pm\sigma^L$, whose interface with $\sigma^L$ (i.e., the set of bonds
satisfied in $\sigma'^L$ but not in $\sigma^L$, or vice-versa) had the
property that the number of interface bonds scaled as $L^{d_s}$, with
$d_s<d$.

Palassini and Young (PY) \cite{PY00}, working in both $d=3$ and $4$,
excited $\sigma^L$ differently by adding a novel coupling-dependent bulk
perturbation to the Hamiltonian (\ref{eq:Hamiltonian}) in $\Lambda_L$.  PY
also interpreted their results as evidence of an excited $\sigma'^L$ with
$d_s<d$.  At the same time the interface energy scaled as $l^{\theta'}$
($\sim L^{\theta'}$), with $\theta'=0$ \cite{theta}; i.e., it remains
finite as $l\to\infty$ (presuming that $l$ continues to grow as $O(L)$).

It is generally agreed \cite{KPY} that the excitations of KM and of PY
correspond to the same physical objects.  We hereafter refer to them as
{\it KMPY excitations\/}.  Given a ground state $\sigma^L$ (or ground state
pair $\pm\sigma^L$) in $\Lambda_L$ with coupling-independent boundary
conditions (such as periodic), KMPY excitations are global spin excitations
whose structural and energetic properties scale with their length in a
unique manner.  More precisely, they are excitations $\sigma'^L$ that are
characterized by three properties: 1) they have $O(L^d)$ spins flipped from
$\pm\sigma^L$ (and thus also have lengthscale $l=O(L)$); 2) the dimension
$d_s$ of their interface with $\sigma^L$ satisfies $d_s<d$; and 3) their
energy difference with $\sigma^L$ (i.e., the interface energy) scales as
$l^{\theta'}$ with $\theta'=0$; i.e., the excitation energy remains of
order one independently of $l$ (and $L$).

Questions have been raised over the correct interpretation of these
numerical results \cite{Middleton00}, and correspondingly whether KMPY
excitations really exist in the spin glass phase \cite{MP00a,MP00b}.
We do not address these questions here.  Rather, we take the point of
view that if KMPY excitations {\it do\/} exist, then their physical
meaning and implications for 
the low-temperature spin glass phase could be 
fundamentally important in the physics of disordered systems.

However, given the recentness of the discovery of KMPY excitations, their
physical meaning and relevance remain unclear.  By showing here rigorously
that their interfaces {\it cannot\/} be pinned by the quenched disorder, we
conclude that they cannot yield new ground or pure states. 
Such restrictions on their large-scale structure clarify the
physical role they might play at low temperature $T$ and provide
general results on a type of ground and/or pure state multiplicity 
not heretofore investigated: the possibility in spin glasses of {\it
regional congruence\/} \cite{HF87a} (see below).

{\it Pinning.}  A crucial question about a new type of
excitation is whether its boundary is {\it pinned\/}.  This 
has not yet been addressed for KMPY interfaces, to our
knowledge.

To understand pinning in this context, consider at $T=0$ an increasing
sequence of $\Lambda_L$'s.  For each $L$, use the procedures of
\cite{KM00} or \cite{PY00} to create KMPY excitations; e.g. \cite{PY00},
first generate the periodic boundary condition ground state $\sigma^L$, and
then add a bulk perturbation (Eq.~(2) of \cite{PY00}) to the Hamiltonian to
generate $\sigma'^{L}$, a perturbed ground state (i.e., an excited state
for the original Hamiltonian).  Then study the bonds $\langle x,y\rangle$
that obey
\begin{equation}
\label{eq:interface}
\sigma^L_x\sigma^L_y=-\sigma'^L_x\sigma'^L_y \ , 
\end{equation}
i.e., those that are satisfied in one state but not the other.  The
corresponding set of bonds in the dual lattice comprises the finite-size
``domain wall'' (for that $L$) or interface of the excitation (i.e., the
boundary of the set of spins that are flipped to go from $\sigma^L$ to
$\sigma'^L$).

By pinning we mean the following.  Consider a fixed $L_0$ (which can be
arbitrarily large).  Apply the KM or PY procedure on cubes $\Lambda_L$,
with $L\gg L_0$.  Observe $\sigma'^{(L,L_0)}$, the excited spin
configuration $\sigma'^L$ restricted to $\Lambda_{L_0}$.  If the excitation
interface remains inside $\Lambda_{L_0}$ as $L\to\infty$
\cite{convergence}, then the interface is {\it pinned\/}. 
Pinning of lower-dimensional interfaces by quenched disorder is
known to occur, e.g., in disordered ferromagnets 
for sufficiently large $d$ \cite{HH85,K85,BK94}; this
example, and its relevance to spin glasses, will be discussed further below.

If the interface is {\it not\/} pinned, we say it ``deflects to
infinity''.  Here, for any fixed $L_0$, the
interface, for all $L$ above some $L'$, will {\it not\/}
enter $\Lambda_{L_0}$.  (This is what occurs with
interface ground states in disordered ferromagnets for small $d$
\cite{HH85,K85,BK94}.)  See Fig.~1 for a schematic illustration.

\begin{figure}
\vspace{-0.4in}
\centerline{\epsfig{file=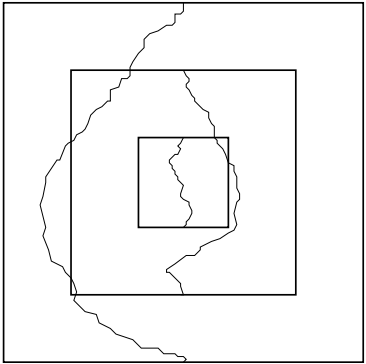,width=2.0in}}
\vspace{-0.2in}
\caption{A sketch of interface deflection to infinity for $d=2$. As 
$L$ increases, the interface recedes from the origin.  
The interfaces eventually are completely outside
any fixed square.  (The deflection can scale more slowly with $L$
than in the figure.)}
\label{fig:deflect}
\end{figure}

{\it Scenarios.\/} Assuming now that KMPY excitations do exist on at
least small to moderate lengthscales, there are four possibilities for
larger lengthscales.  One is that the
excitations disappear altogether, noted as a
possibility in \cite{PY00}.  Of the three remaining (more
interesting) scenarios, two have the KMPY interface pinned,
and the third has it deflect to infinity.

Scenario (1): KMPY excitations are pinned and give rise to new ground
states at $T=0$ and new pure states at $T>0$ (below some $T_c$).  I.e.,
their interfaces with the original ground state, or with each other, become
relative domain walls between distinct ground (or pure) states not related
by a global spin flip.  (For more details on relative domain walls, see
\cite{NS00,NS01}.)  This interesting picture differs from previous
proposals for the low-$T$ spin glass phase, and would be the first example
of {\it regional congruence\/} \cite{HF87a} in spin glasses.  Two distinct
thermodynamic states that are not global flips of each other are regionally
congruent if the total relative domain wall density vanishes; i.e.,
essentially the condition above that $d_s<d$.  (If the density is nonzero
so that $d_s=d$, as is more usually supposed, the states are said to be
{\it incongruent}.)

Scenario (2): The excitations are pinned, but do {\it not\/} give rise
to new ground states.  For this to occur, any $L\to\infty$
limit of $\sigma'^L$ must be energetically unstable (in the 
Hamiltonian (\ref{eq:Hamiltonian})) to the flip of some fixed {\it
finite\/} droplet.  Even if this occurs,
it could be that KMPY excitations, at
$T>0$, still give rise to new {\it pure\/} states.

Scenario (3): KMPY excitations persist on all lengthscales, but their
interfaces deflect to infinity, as in Fig.~1.  If this occurs, they cannot
give rise to new thermodynamic ground or pure states, but could still be
relevant to the excitation spectrum in finite volumes.

Which of these scenarios actually occurs was not addressed in
\cite{PY00}, although \cite{HKM} as well as \cite{HHD} implicitly or
explicitly assume Scenario (1).  Determining which 
occurs is crucial to understanding the role of
these excitations in the spin glass phase.

We now prove that {\it KMPY excitations cannot be pinned by the quenched
disorder, ruling out Scenarios (1) and (2)\/}.  (A heuristic argument
against Scenario (1) was presented in \cite{NSComment}.)  The remaining
possibility (in which KMPY excitations persist on all lengthscales) is
Scenario (3), where they deflect to infinity.  Before presenting our
theorem, we introduce the concept of {\it metastate\/}, which is implicit
in the theorem and useful for understanding its applications.

{\it Metastates.\/} A ($T=0$) metastate gives the probability of finding 
various ground state pairs in typical large volumes.  There are different
constructions (technical details are in
\cite{NS00,AW,NS96b,NS97}); we give here only a simple physical
description.

Consider the cube $\Lambda_{L_0}$ with (large) fixed $L_0$, and then
examine a sequence of $\Lambda_L$'s with $L\to\infty$, all with (for
example) periodic boundary conditions.  Look, for each $L$, at the part of
the ground state $\sigma^L$ inside the smaller fixed $\Lambda_{L_0}$, and
keep a record of the fractions of $L$'s in which different spin
configurations appear inside $\Lambda_{L_0}$.  (If there's only a single
pair of thermodynamic ground states, then one eventually sees one fixed
configuration inside $\Lambda_{L_0}$, and thus asymptotically the ground
state pair appears inside $\Lambda_{L_0}$ for a fraction one of the $L$'s).
The resulting ($T=0$, periodic boundary condition) metastate is a
probability measure on infinite-volume ground state pairs that provides
information on the fraction of $L$'s for which each of them appears inside
$\Lambda_{L_0}$.
 
For the coming theorem, we extend the notion of metastate to {\it uniform
perturbation metastate\/} \cite{excitation}.  For each $L$, consider, in
addition to $\sigma^L$, a second $\sigma'^L$ (which in practice will be the
ground state of a perturbed Hamiltonian) chosen so as to respect the
finite-volume (torus) translation-invariance already present for $({\cal
J}^L,\sigma^L)$ due to the periodic boundary conditions.  (We show later
that this holds for $\sigma'^L$ in both KM and PY constructions.)  Then do
for the {\it pair\/} $(\sigma^L,\sigma'^L)$ what was done for $\sigma^L$ in
the original metastate.  The resulting uniform perturbation metastate gives
for both (infinite-volume) ground and excited states their relative
frequency of appearance inside large volumes.  The theorem presents
implications of translation-invariance on the resulting interfaces, and
shows that if there is a (pinned) interface at all, it must have
strictly positive density (so $d_s=d$).

{\it Theorem.\/} On the cube $\Lambda_L$ with periodic boundary
conditions, for a given coupling configuration ${\cal J}^L$ let
$\sigma^L$ and $\sigma'^L$ be a pair of spin configurations such that
the joint distribution of $({\cal J}^L,\sigma^L,\sigma'^L)$ is
invariant under (torus) translations of $\Lambda_L$.  Let $({\cal
J},\sigma,\sigma')$ be any limit in distribution as $L\to\infty$ of
$({\cal J}^L,\sigma^L,\sigma'^L)$. Then for almost every $({\cal
J},\sigma,\sigma')$ either $\sigma'=\pm\sigma$ or else $\sigma$ and
$\sigma'$ have a relative interface of strictly positive density.

{\it Proof.\/} Because the joint distribution $\kappa^L$ of $({\cal
J}^L,\sigma^L,\sigma'^L)$ is for every $L$ invariant under torus
translations, any limiting distribution $\kappa$ of $({\cal
J},\sigma,\sigma')$ is invariant under all translations of the
infinite-volume cubic lattice.  The
translation-invariance of $\kappa$ allows its decomposition into
components in which translation-ergodicity holds (see,
e.g., \cite{NS97,NS95}).  For each bond $\langle x,y\rangle$ consider
the event $A_{\langle x,y\rangle}$ that $\langle x,y\rangle$ is
satisfied in one but not the other of $\sigma,\sigma'$.  In each
ergodic component, either the probability of $A_{\langle x,y\rangle}$
equals zero and then there is no interface (i.e.,
$\sigma'=\pm\sigma$) or else it equals some $\rho>0$.  In the
latter case, by the spatial ergodic theorem (see, e.g.,
\cite{NS97,NS95}) the spatial density of $\langle x,y\rangle$'s such
that $A_{\langle x,y\rangle}$ occurs must equal $\rho$, i.e., the
$(\sigma,\sigma')$ interface has a nonzero density.  Thus there is
zero probability (with respect to $\kappa$) of 
a $({\cal J},\sigma,\sigma')$ such that there is a $\sigma,\sigma'$
interface, but with zero density.

\medskip

{\it Remarks.\/} 1) If one takes the $\kappa({\cal J},\sigma,\sigma')$ of
the theorem and conditions it on the coupling realization ${\cal J}$, the
resulting conditional distribution $\kappa_{\cal J}(\sigma,\sigma')$ is the
uniform perturbation metastate discussed above.  Equivalent to the
theorem's conclusion is that for almost every ${\cal J}$, if $\sigma$ and
$\sigma'$ are chosen from $\kappa_{\cal J}(\sigma,\sigma')$, there is zero
$\kappa_{\cal J}$-probability of an (infinite-volume) interface with zero
density.  This rules out $(\sigma^L,\sigma'^L)$ interfaces with $d_s<d$
that are also pinned.

2) Although the theorem as formulated here addresses ground
and excited states at $T=0$, it should be extendable to pure states at
(low) $T>0$ by ``pruning'' small thermally induced droplets.

3) Although our construction used periodic boundary conditions (both for
simplicity and because KM and PY used them), the theorem can be applied to
other boundary conditions chosen independently of the couplings (as in
\cite{NS01}, to which we refer the reader for details).

4)  Although our focus here is on spin glasses, the theorem applies
equally to many other systems, including disordered and homogeneous
ferromagnets.

{\it Application to KMPY excitations.\/} An immediate application of the
theorem is that KMPY excitations cannot be pinned, and so cannot give rise
to (regionally congruent) ground states.  To see that the theorem
applies to the PY construction, note first that the periodic boundary
condition clearly implies torus translation invariance for the distribution
of $({\cal J}^L,\sigma^L)$.  The Hamiltonian perturbation (Eq.~(2) of
\cite{PY00}) is constructed from $\sigma^L$ in a translation-covariant
way, and thus the distribution of $({\cal
J}^L,\sigma^L,\sigma'^L)$ is also invariant.

{\it A priori,\/} the situation for the KM construction is more subtle.
Here if $\sigma'^L$ results from a {\it fixed\/} pair of spins at sites
$z_0,z'_0$, translation-invariance would be lost.  So instead, for fixed
${\cal J}^L$ and $\sigma^L$ we regard $\sigma'^L$ as the {\it random\/}
excitation resulting when $z,z'$ are chosen for each $\Lambda_L$ from the
uniform distribution on its sites; this restores translation-invariance.  A
first result is then that for a fraction one (as $L\to\infty$) of such
choices of $z,z'$, the interface cannot be pinned.  But there's a second,
more noteworthy result: it follows that a {\it fixed\/} $z_0,z'_0$ cannot
yield a KMPY (or similar) excitation (which now would necessarily be
pinned) at all, since if it did there would be a positive fraction of
random $(z,z')$'s yielding the same excitation.  {\it Thus a fixed
$z_0,z'_0$ must either yield a droplet excitation with volume $o(L^d)$
(e.g., bounded) as $L\to\infty$ or else one with $d_s=d$.}

{\it Regional Congruence\/}.  As discussed above, regionally congruent
thermodynamic states \cite{HF87a} are those whose relative interfaces
have zero density.  Examples are the interface states in homogeneous
and disordered Ising ferromagnets for sufficiently large $d$.  
But prior to \cite{KM00} and \cite{PY00} there had been no evidence for the
existence of regional congruence in spin glasses.

In \cite{NS00} and \cite{NS01}, we proved that thermodynamic states
generated by coupling-independent boundary conditions (e.g., periodic,
antiperiodic, free, fixed, etc.) are (with probability one) either the same
(modulo a global spin flip) or else incongruent (so $d_s=d$).  Thus
regionally congruent states can {\it only\/} arise, if at all, through a
sequence of $\Lambda_L$'s with {\it coupling-dependent\/} boundary
conditions; i.e., those that are {\it conditioned\/} on the ${\cal J}^L$'s.
This is consistent with the above theorem, because such boundary conditions
typically violate translation-invariance.  E.g., interface states in
ferromagnets arise through boundary conditions, such as Dobrushin
\cite{Dob,NShd} (i.e., plus boundary spins above the ``equator'' and minus
below), that are not translation-covariant and implicitly use the knowledge
that all couplings are positive.  But in a spin glass, they are now
gauge-equivalent to boundary conditions that {\it are\/} both
translation-covariant and coupling-{\it independent\/}.

But now conclusions beyond those of \cite{NS00,NS01} follow from the
theorem, because in addition regional congruence cannot arise
through any translation-covariant construction.  A consequence is that any
algorithm of the types currently known {\it cannot\/} yield regionally
congruent states.  The KM and PY procedures typify two main approaches in
the search for new excitations and/or states, given that {\it a priori\/}
the interface location is unknown: either a uniformly random sampling
procedure (KM) or else a global perturbation (PY) (or similar coupling to a
carefully chosen external field).  Although the latter approach {\it is\/}
coupling-dependent, it is also translation-covariant and so cannot generate
regional congruence.  One method that {\it could\/} find regional
congruence, if it occurs, is an exhaustive search through all
$2^{O(L^{d-1})}$ fixed boundary conditions on $\Lambda_L$ for each $L$; but
that is hardly an option.

We conclude by noting that the order one energy difference of
$\sigma,\sigma'$ plays no role in our theorem, which in fact has far wider
applicability.  Only {\it one\/} of its applications is to KMPY
excitations, but these have generated interest {\it because\/} both
$\theta'=0$ and $d_s<d$.  However, the order one energy does play a crucial
role in our earlier heuristic argument \cite{NSComment}, which we briefly
summarize (and slightly extend) here.  It was conjectured \cite{FH86} and
subsequently proved \cite{AFunpub} (also \cite{NS92}) that free energy
fluctuations in spin glasses with coupling-independent boundary conditions
on $\Lambda_L$ scale as $L^\theta$, with $\theta\le (d-1)/2$.  It should
then follow that the exponent $\theta_r$ characterizing interface energies
for regionally congruent states satisfies the bound $\theta_r>\theta$ (as
argued in \cite{NS92,FH87b,vE}) because otherwise regional congruence could
be seen with coupling-independent boundary conditions.  So pinned
interfaces with $d_s<d$ (hence regionally congruent states) must have
$\theta_r>\theta$; if the latter inequality is violated, as in KMPY
excitations, then the interface should deflect to infinity.

{\it Acknowledgments.\/} This research was supported in part by NSF
Grants DMS-98-02310 (CMN) and DMS-98-02153 (DLS).  The authors thank
Peter Young and Matteo Palassini for useful correspondence.  They also
thank the Santa Fe Institute, where part of this work was done, for
its hospitality.


\begin{references}

\bibitem{KM00}
F.~Krzakala and O.C.~Martin, Phys.~Rev.~Lett.~{\bf 85}, 3013 (2000).
Also available as cond-mat/0002055.

\bibitem{PY00}
M.~Palassini and A.P.~Young, Phys.~Rev.~Lett.~{\bf 85}, 3017 (2000).
Also available as cond-mat/0002134.

\bibitem{HKM}
J.~Houdayer, F.~Krzakala, and O.C.~Martin,
available as cond-mat/0009382.

\bibitem{KPY}
H.G.~Katzgraber, M.~Palassini, and A.P.~Young,
available as cond-mat/0007113.

\bibitem{Middleton00}
A.A.~Middleton,
Phys.~Rev.~B~{\bf 63}, 060202(R) (2001),
available as cond-mat/0007375.

\bibitem{MP00a} 
E.~Marinari and G.~Parisi, 
Phys.~Rev.~B~{\bf 62}, 11677 (2000),
available as cond-mat/0005047.

\bibitem{MP00b}
E.~Marinari and G.~Parisi, 
available as cond-mat/0007493.

\bibitem{HHD}
G.~Hed, A.K.~Hartmann, and E.~Domany,
available as cond-mat/0012451.

\bibitem{EA}
S.F.~Edwards and P.W.~Anderson
J.~Phys.~F {\bf 5}, 965 (1975).

\bibitem{theta} This is different from the exponent $\theta$
governing the energy change between coupling-independent boundary
conditions, such as periodic to antiperiodic.  Numerical studies in $d=3,4$
suggest $\theta>0$ (see, e.g., D.S.~Fisher and D.A.~Huse, Phys.~Rev.~B {\bf
38}, 386 (1988)).

\bibitem{HF87a}
D.A.~Huse and D.S.~Fisher,
J.~Phys.~A {\bf 20}, L997 (1987).

\bibitem{convergence} We require this only for {\it some\/}
arbitrarily large $L$'s, but presumably it will then be true for all
large $L$ with the interface converging to a well-defined limit as
$L\to\infty$.

\bibitem{HH85}
D.A.~Huse and C.L.~Henley,
Phys.~Rev.~Lett.~{\bf 54}, 2708 (1985).

\bibitem{K85}
M.~Kardar,
Phys.~Rev.~Lett.~{\bf 55}, 2923 (1985).

\bibitem{BK94}
A.~Bovier and C.~K\"{u}lske,
Rev.~Math.~Phys.~{\bf 6}, 413 (1994).

\bibitem{NS00}
C.M.~Newman and D.L.~Stein,
Phys.~Rev.~Lett.~{\bf 84}, 3966 (2000).

\bibitem{NS01}
C.M.~Newman and D.L.~Stein, 
available as cond-mat/0103395.

\bibitem{NSComment}
C.M.~Newman and D.L.~Stein, 
available as cond-mat/0010033.

\bibitem{AW}
M.~Aizenman and J.~Wehr,
Commun.~Math.~Phys.~{\bf 130}, 489 (1990).

\bibitem{NS96b}  
C.M.~Newman and D.L.~Stein,
Phys.~Rev.~Lett.~{\bf 76}, 4821 (1996).

\bibitem{NS97}
C.M.~Newman and D.L.~Stein,
Phys.~Rev.~E {\bf 55}, 5194 (1997).

\bibitem{excitation} This is related to the {\it excitation
metastate\/} of \cite{NS01}, but with the
difference that here perturbations are
applied in a uniform, translation-invariant manner.

\bibitem{NS95}
C.M.~Newman and D.L.~Stein,
Phys.~Rev.~Lett.~{\bf 76}, 515 (1996).

\bibitem{Dob}
R.L.~Dobrushin,
Theor.~Prob.~Appl.~{\bf 17}, 582 (1972).

\bibitem{NShd}
C.M.~Newman and D.L.~Stein, 
Phys.~Rev.~E~{\bf 63}, 16101-1 (2001).

\bibitem{FH86}
D.S.~Fisher and D.A.~Huse,
Phys.~Rev.~Lett.~{\bf 56}, 1601 (1986).

\bibitem{AFunpub}
M.~Aizenman and D.S.~Fisher, unpublished.

\bibitem{NS92}
C.M.~Newman and D.L.~Stein,
Phys.~Rev.~B~{\bf 46}, 973 (1992).

\bibitem{FH87b}
D.S.~Fisher and D.A.~Huse,
J.~Phys.~A {\bf 20}, L1005 (1987).

\bibitem{vE}
A.~van~Enter, J.~Stat.~Phys.~{\bf 60}, 275 (1990).

\end{references}
\end{document}